\documentstyle[natbib209]{agile7th}
\input psfig.sty

\def \AAP #1 #2 {{\em Astron. Astrophys.\/} {\bf #1}, #2}
\def \AAL #1 #2 {{\em Astron. Astrophys. Lett.\/} {\bf #1}, L#2}
\def \AAR #1 #2 {{\em Astron. Astrophys. Rev.\/} {\bf #1}, #2}
\def \AAS #1 #2 {{\em Astron. Astrophys. Suppl. Ser.\/} {\bf #1}, #2}
\def \AJ #1 #2 {{\em Astron. J.\/} {\bf #1}, #2}
\def \ANNREV #1 #2 {{\em Ann. Rev. Astron. Astrophys.\/} {\bf #1}, #2}
\def \APJ #1 #2 {{\em Astrophys. J.\/} {\bf #1}, #2}
\def \APJL #1 #2 {{\em Astrophys. J. Lett.\/} {\bf #1}, L#2}
\def \APJS #1 #2 {{\em Astrophys. J. Suppl.\/} {\bf #1}, #2}
\def \APSS #1 #2 {{\em Astrophys. Space Sci.\/} {\bf #1}, #2}
\def \ASR #1 #2 {{\em Adv. Space Res.\/} {\bf #1}, #2}
\def \BAIC #1 #2 {{\em Bull. Astron. Inst. Czechosl.\/} {\bf #1}, #2}
\def \JSQRT #1 #2 {{\em J. Quant. Spectrosc. Radiat. Transfer\/} {\bf #1}, #2}
\def \MN #1 #2 {{\em Mon. Not. R. Astr. Soc.\/} {\bf #1}, #2}
\def \MEM #1 #2 {{\em Mem. R. Astr. Soc.\/} {\bf #1}, #2}
\def \PLR #1 #2 {{\em Phys. Lett. Rev.\/} {\bf #1}, #2}
\def \PASJ #1 #2 {{\em Publ. Astron. Soc. Japan\/} {\bf #1}, #2}
\def \PASP #1 #2 {{\em Publ. Astr. Soc. Pacific\/} {\bf #1}, #2}
\def \NAT #1 #2 {{\em Nature\/} {\bf #1}, #2}
\def \SAIT #1 #2 {{\em Mem.\ Soc.\ Astron.\ It.\/} {\bf #1}, #2}
\def \MESS #1 #2 {{\em The Messenger\/} {\bf #1}, #2}
\def \ASTRNACH #1 #2 {{\em Astron. Nach.\/} {\bf #1}, #2}
\def \AGPSR #1 #2 {{\em ASI Special Publication\/} {\bf #1}, #2}

\begin{opening}

\title{Microquasars in the GeV-TeV era}
\author{J.M. Paredes, V. Zabalza}
\institute{Departament d'Astronomia i Meteorologia and Institut de Ci\`encies
del Cosmos (ICC), Universitat de Barcelona (UB/IEEC)}
\date{} 
\end{opening}

\begin{document}

\oddpagefooter{}{}{} 
\evenpagefooter{}{}{} 
\medskip  

\begin{abstract} 
    The discovery of non-thermal X-ray emission from the jets of some X-ray
    binaries, and especially the discovery of GeV-TeV gamma-rays in some of
    them, provide a clear evidence of very efficient acceleration of particles
    to multi-TeV energies in these systems. The observations demonstrate the
    richness of non-thermal phenomena in compact galactic objects containing
    relativistic outflows or winds produced near black holes and neutron stars.
    We review here some of the main observational results on the non-thermal
    emission from X-ray binaries as well as some of the proposed scenarios to
    explain the production of high-energy gamma-rays. 
\end{abstract}

\medskip

\section{X-ray binaries and microquasars}

X-ray binaries (XB) are binary systems containing a compact object (a neutron star
or a stellar mass black hole) accreting matter from the companion star.
Depending on the spectral type of the optical companion they are classified in
High Mass X-ray binaries (HMXB) or Low Mass X-ray Binaries. For HMXBs the
optical companion has an early type (either O or B) and mass transfer takes
place via a decretion disc for Be stars, wind accretion or Roche lobe overflow.
LMXBs, on the other hand, have an optical companion with spectral type later
than B and mass transfer takes place through Roche lobe overflow. Up to now a
total of 299 XBs have been catalogued, out of which 114 are HMXBs and 185 are
LMXBs \citep{liu06,liu07}. 

Microquasars are X-ray binaries that have relativistic jets resulting from the
accretion onto the compact object. These relativistic jets can emit non-thermal
radio emission through synchrotron radiation, so all radio emitting XBs (REXB)
are microquasar candidates. 65 REXBs have been detected (22\% of all XBs): 9 of
them are HMXBs and 56 are LMXBs. The interaction of the relativistic jets with
the optical companion radiation or with the interstellar medium can result in
GeV-TeV emission from these sources. Below we explore the observations of a few
selected microquasars that indicate the presence of accelerated non-thermal
particle population susceptible to emit in the gamma-ray domain.

\subsection{Superluminal jets: GRS~1915+105}

The first clear evidence of relativistic jets in XRBs was found by
\cite{mirabel94} through the detection of superluminal motion in the ejecta of
the microquasar GRS 1915+105. It was proposed that relativistic electron
population in the jet could emit above MeV  energies through Inverse Compton
scattering or even direct synchrotron \citep{atoyan99}. However, this particular
source hasn't been detected yet in neither the GeV nor the TeV band
\citep{saito09,szostek09}. 

\subsection{Strong radio outbursts: Cygnus~X-3}

Other galactic accreting sources have been studied best through their radio
outbursts, as is the case of Cygnus~X-3. This HMXB is formed by a Wolf Rayet
star and a compact object that is thought to be a neutron star for orbit
inclination angles above 60$^\circ$ or a black hole otherwise \citep{vilhu09}. 
Cyg~X-3 shows flaring levels of up to 20~Jy, and was first detected and closely
observed at this level in 1972, resulting in one of the best-known examples of
expanding synchrotron emitting sources. These outbursts can be modeled
successfully as coming from particle injection in twin jets \citep{marti92},
which have been subsequently imaged trough interferometric techniques
\citep{marti01,miller04}.

Long-term multiwavelength monitoring of Cyg~X-3 has revealed that strong
radio flares occur only when the source shows high soft X-ray flux and
a hard power-law tail. If the electrons responsible for the strong radio
outbursts and the hard X-ray tails are accelerated to high enough energies,
detectable emission in the $\gamma$-ray energy band is possible. In the last few
months, both the AGILE/GRID \citep{tavani09} and Fermi/LAT \citep{abdo09c}
collaborations have published clear detections of Cyg~X-3 in high energy
gamma-rays.

\subsection{Jet-medium interaction: SS~433}

For other microquasars, one of the most prominent features is the interaction
between their relativistic jet and the interstellar medium surrounding them.
While theoretical predictions for gamma-ray emission have been made
\citep{bordas09}, none has been detected yet. The most notable case with this
feature is SS~433, a HMXB with twin relativistic precessing jets. The precession
has been clearly observed in the radio domain below arcsecond scales
\citep{stirling02}. At a larger scale, the interaction of the jets with the
surrounding parent nebula W50 has deformed the originally spherical nebula into a
twisting elongated shape \citep{dubner98}. This source is the only one for which
the jets are known to contain a hadronic component after Doppler shifted iron
lines were detected in spatially resolved regions corresponding to the jet and
counter-jet, proving that particle re-acceleration in relativistic jets does not
only affect electron but also atomic nuclei \citep{migliari02}.

\section{Detected Binary TeV sources}

Among the VHE sources detected with the \v{C}erenkov telescopes there are three
clearly associated to X-ray binaries. These Binary TeV sources (BTV),
PSR~B1259$-$63 \citep{aharonian05a}, LS~I~+61~303 \citep{albert06} and LS~5039
\citep{aharonian05b}, have a bright high-mass primary star, which provides an
intense UV seed photon field for inverse Compton scattering of particles
accelerated around the compact object. All of them have been detected at TeV
energies in several parts of their orbits and show variable emission and hard
spectrum. The emission is periodic in the systems LS~5039 and LS~I~+61~303, with
a period of 3.9078$\pm$0.0015 days and 26.8$\pm$0.2 days respectively
\citep{aharonian06, albert09},  consistent with their orbital periods
\citep{casares05a, casares05b, aragona09}.  These two sources also share the
distinction of being the only two known high-energy emitting X-ray binaries that
are spatially coincident with sources above 100~MeV listed in the Third EGRET
catalog \citep{hartman99}. LS~5039 is associated with 3EG~J1824$-$1514
\citep{paredes00} and LS~I~+61~303 with 3EG~J0241+6103 \citep{kniffen97}. Both
sources have also been detected by the Fermi observatory
\citep{abdo09a,abdo09b}.  In the case of PSR~B1259$-$63 the periodicity has not
been yet determined because the long orbital period (3.4 years) requires
extensive monitoring during several years with the \v{C}erenkov telescopes. The
source was not detected by EGRET.

Another high-mass X-ray binary, Cygnus~X-1, was observed with MAGIC during a
short-lived flaring episode, and strong evidence (4.1$\sigma$ post-trial
significance) of TeV emission  was found \citep{albert07}. 

The nature of the compact object is well determined in only two sources. In the
case of Cygnus~X-1 it is a black hole and, in the case of PSR~B1259$-$63, a
neutron star. For LS~I~+61~303 and LS~5039 there is no strong evidence yet
supporting either the black hole nor the neutron star nature of the compact
objects. Some of these sources have also been detected at MeV and GeV energies
by instruments onboard the Compton Gamma-ray Observatory (CGRO), like COMPTEL
and EGRET (as commented above), or by the two current spatial missions ${\it
AGILE}$ and ${\it Fermi}$.  A common characteristic of these four sources is
that all of them are radio emitters, producing non-thermal radiation. These
sources, with the exception of PSR~B1259$-$63, show elongated radio structures
of synchrotron origin. It is possible that PSR~B1259$-$63 has this kind of
structure but has not yet been detected with the present sensitivity and
resolution of the instruments available in the southern hemisphere.


\subsection{PSR~B1259$-$63}

PSR~B1259$-$63 is the first variable galactic source of VHE gamma-rays
discovered. It is a binary system containing a Be main sequence donor, known
as LS~2883, and a 47.7~ms radio pulsar orbiting it every 3.4 years in a very
eccentric orbit with $e=0.87$ \citep{johnston94}. 

The radiation mechanisms and  interaction geometry in this pulsar/Be star system
were studied in \cite{tavani97}. In a hadronic scenario, the TeV  light-curve,
and radio/X-ray light-curves, can be produced by the collisions of high energy
protons accelerated by the pulsar wind and the circumstellar disk
\citep{neronov07}. A very different model is presented in \cite{khangulyan07},
where it is shown that the TeV light curve can also be explained by IC scenarios
of gamma-ray production.

\subsection{LS~I~+61~303}

This source is located at a distance of 2.0$\pm$0.2~kpc \citep{frail91}. It
contains a rapidly rotating B0~Ve star with a stable equatorial shell, and a
compact object of unknown nature with a mass between 1 and 5 $M_\odot$, orbiting
it every 26.5~d \citep{hutchings81, casares05a}. Quasi-periodic radio outbursts
monitored during 23 years have provided an accurate orbital period value of
$P_{\rm orb}$=26.4960$\pm$0.0028~d and the presence of a 4.4 yr superorbital
periodicity in the phase location and amplitude of the outburst
\citep{paredes87,gregory02}. The orbit is eccentric ($e\simeq0.72$) and
periastron takes place at phase 0.23$\pm$0.02, assuming T$_0$= JD 2,443,366.775
\citep{casares05a}. \cite{grundstrom07} obtained new orbital parameters,
revealing an eccentricity of 0.55 and a periastron at phase 0.30$\pm$0.01.
However, the Balmer lines used in their study are possibly contaminated by
the stellar wind.  New radial velocities measurements reported recently by
\citet{aragona09} give improved orbital elements for LS~I~+61~303 and for
LS~5039.

An orbital X-ray periodicity has been found using {\it RXTE}/ASM archival data
\citep{paredes97}.  Simultaneous X-ray ({\it RXTE}/PCA) and radio observations
of LS~I~+61~303 over the 26.5 day orbit showed a significant offset between the
peak of the X-ray and radio flux. The X-ray outbursts, starting around phase 0.4
and lasting up to phase 0.6, peak almost half an orbit before the radio ones
(\cite{harrison00} and references therein).  Similar results have recently been
obtained at higher energies with {\it INTEGRAL} data \citep{hermsen06}. Recently
the MAGIC collaboration have detected correlated X-ray and VHE emission from
LS~I~+61~303 during a simultaneous multiwavelength campain covering 60\% of an
orbit \citep{anderhub09}, suggesting that the same leptonic particle population
is responsible for the emission in these two energy bands.  The maximum of the
radio outbursts varies between phase 0.45 and 0.95.  \cite{massi04} reported the
discovery of an extended jet-like and apparently precessing radio emitting
structure at angular extensions of 10--50~milliarcseconds. VLBA images obtained
during a full orbital cycle show a rotating elongated morphology
\citep{dhawan06}, which may be consistent with a model based on the interaction
between the relativistic wind of a young non-accreting pulsar and the wind of
the stellar companion (\cite{dubus06}; see nevertheless  \cite{romero07} for a
critical discussion of this scenario). 

\begin{figure}[ht!]
\centerline{\psfig{figure=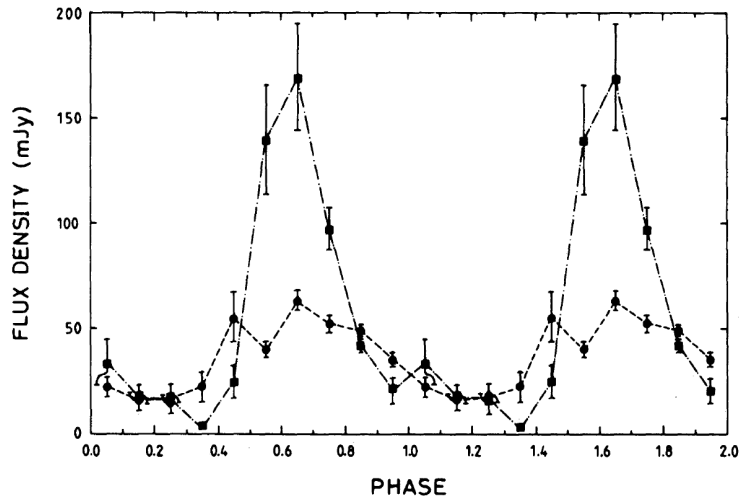,height=6.92cm,width=10cm,angle=0}}
\caption{Average radio lightcurves for the active phase (squares, dash-dotted)
and quiet phase (circles, dashed) of the 4.4~yr superorbital period. Figure
adapted from Paredes et al.~(1990). It will be very interesting to explore
wether this modulation is also present in the HE gamma-ray domain. \label{fig:lsi_radio}}
\end{figure}

The gamma-ray lightcurve observed by {\it Fermi} and reported by \cite{abdo09b}
shows hints of an orbit-to-orbit variability. The increase in integrated orbital
flux observed during the past months could be related to the 4.4 yr superorbital
periodicity. Given the similarity of the gamma-ray and radio orbital lightcurves
of the source, a similar long term behavior is to be expected. The analysis of
{\it Fermi} data during the following months will allow us to better understand
the multiple components that conform the broad band SED of this peculiar source.
In Fig.~\ref{fig:lsi_radio} we show the difference in the radio peak amplitude
between an active phase and a quite phase of the superorbital period.

\subsection{LS~5039}

LS 5039 is a high-mass X-ray binary containing a compact object of unknown mass,
1.5--10 $M_\odot$ depending on the binary system inclination, which is an
unknown parameter. The neutron star or black hole orbits an O6.5V((f)) star
every 3.9 days in an eccentric orbit with e = 0.35, and is located at 2.5~kpc
\citep{casares05b}. The radio emission of LS~5039 is persistent, non-thermal and
variable, but no strong radio outbursts or periodic variability have been
detected so far \citep{ribo99, ribo02}. VLBA observations allowed for the
detection of an elongated radio structure, interpreted as relativistic jets
\citep{paredes00}. The discovery of this bipolar radio structure, and the fact
that LS~5039 was the only source in the field of the EGRET source
3EG~J1824$-$1514 showing X-ray and radio emission, allowed to propose the
physical association of both sources \citep{paredes00}. A theoretical discussion
of the radio properties of LS~5039 can be found in \citet{bosch09}.

X-ray observations of LS~5039 with the {\it Suzaku} satellite, covering one and
half orbits, show strong modulation over the orbital period of the system that
is closely correlated with the TeV gamma-ray light curve \citep{takahashi09}.
Furthermore, when comparing this lightcurve with observations taken more than
ten years ago it has been found to be extremely stable over long periods
\cite{kishishita09}.  The X-ray/TeV correlation seems to indicate a synchrotron
origin for the X-rays, and that the electrons producing the synchrotron
radiation are also responsible for the TeV emission via IC scattering. However,
whereas the TeV periodicity is mainly explained by the photon-photon pair
production and anisotropic IC scattering, the X-ray modulation seems to be
produced by adiabatic losses dominating the synchrotron and IC losses of
electrons \citep{takahashi09}. To avoid heavy absorption, it is required that
the gamma-ray emission be produced at the periphery of the binary system
\citep{khangulyan08,bosch08}, although a scenario accounting for electromagnetic
cascading has been also considered \citep{sierpowska07, khangulyan08}.

Two incompatible scenarios have been proposed to explain the acceleration
mechanism that powers the relativistic electrons. In the first one electrons are
accelerated in the jet of a microquasar powered by accretion \citep{paredes06}.
In the second one they are accelerated in the shock between the relativistic
wind of a young non-accreting pulsar and the wind of the stellar companion
\citep{dubus06}. 

\subsection{Cygnus~X-1}

Cygnus~X-1 is the brightest persistent HMXB in the Galaxy, radiating a maximum
X-ray luminosity of a few times $10^{37}$ erg s$^{-1}$ in the 1--10 keV range.
At radio wavelengths the source displays a $\sim$15~mJy flux density and a flat
spectrum, as expected for a relativistic compact (and one-sided) jet ($v>0.6c$)
during the low/hard state \citep{stirling01}. A transient relativistic radio jet
was observed during a phase of repeated X-ray spectral transitions in an epoch
with the softest 1.5--12 keV X-ray spectrum \citep{fender06}.  Arc-minute
extended radio emission around Cygnus~X-1 was found using the VLA
\citep{marti96}. Its appearence was that of an elliptical ring-like shell with
Cygnus~X-1 offset from the center. Later, as reported in \cite{gallo05}, such
structure was recognised as a jet-blown ring around Cygnus~X-1. This ring could
be the result of a strong shock that develops at the location where the pressure
exerted by the collimated jet, detected at milliarcsec scales, is balanced by
the ISM. The observed thermal Bremsstrahlung radiation would be produced by the
ionized gas behind the bow shock. 

MAGIC very high energy gamma ray observations of this source revealed strong
evidence (at a 4.1$\sigma$ post-trial confidence level) for a short flaring
episode from Cygnus~X-1.  These TeV measurements were coincident with an intense
state of hard X-ray emission observed by {\it INTEGRAL}, although no obvious
correlation between the X-ray and TeV emission was found \citep{malzac08}.  The
detection occurred around the superior conjunction of the compact object, when
the highest gamma-ray opacities are expected. After computing the absorbed
luminosity that is caused by pair creation for different emitter positions, it
has been suggested  that the TeV emitter is located at the border of the binary
system \citep{bosch08}. A recent study of the opacity and acceleration models
for the TeV flare can explain qualitatively the observed TeV spectrum, but not
its exact shape \citep{zdziarski09}. 

\subsection{A new BTV candidate: HESS~J0632+057}

HESS~J0632+057 was discovered by the HESS telescope array as a point-like source
in Monoceros \citep{aharonian07}. Its energy spectrum is consistent with a
power-law with photon index of 2.53 and flux normalisation of $9.1\times
10^{-13}$~cm$^{-2}$~s$^{-1}$~TeV$^{-1}$. No evidence of flux variability was
found. Three different sources - the {\it ROSAT} source 1RXS~J063258.3+054857,
the EGRET source 3EG~J0634+0521 and the star MWC~148 - were suggested as
possible associations with HESS~J0632+057. 

Later X-ray observations with {\it XMM-Newton} revealed a variable X-ray source,
XMMU J063259.3+054801, which is positionally coincident with the massive B0pe
spectral type star MWC~148 (HD~259440) and compatible in position with
HESS~J0632+05 \citep{hinton09}. The X-ray spectrum is hard, and can be fitted
with an absorbed power-law model with a 1 keV normalization of $(5.4\pm
0.4)\times 10^{-5}$~keV$^{-1}$~cm$^{-2}$~s$^{-1}$, a photon index of
1.26$\pm$0.04 and a column density of $(3.1\pm 0.3)\times 10^{21}$~cm$^{-2}$.
The spectral energy distribution (SED) of HESS~J0632+057, assuming that the
sources associated at different spectral bands are the real counterparts, looks
similar to that of the TeV binaries LS~I~+61~303 and LS~5039. 

VERITAS observed HESS~J0632+057 during three differents epochs obtaining no
significant evidence for gamma-ray emission \citep{acciari09}.  The HESS
detection and the VERITAS non detection seems to point to a long-term gamma-ray
variability. This seems to happen also in the X-ray band, when comparing the
{\it XMM-Newton} data \citep{hinton09} and {\it Swift} data taken
contemporaneously with VERITAS \citep{acciari09}. The absence of radio emission
in this area, based on the NRAO VLA Sky Survey (NVSS) catalog \citep{condon98},
seemed to indicate that any possible radio counterpart should be either faint
and/or variable. This suspicion has been confirmed very recently. Radio
observations carried out in 2008 with the VLA at 5~GHz and GMRT at 1280~MHz have
found a faint and unresolved source at the position of MWC~148
\citep{skilton09}.  While the radio flux density at the lower frequency is not
variable, there is a significant flux variability on month timescales at 5~GHz.
The TeV variability, as well as the X-ray and the radio variability clearly
associated with MWC~148, gives suport to the idea proposed by \cite{hinton09}
that HESS~J0632+057 is likely a new gamma-ray binary. However, further
observations are necessary to determine the binarity of MWC~148.

\section{Summary}

The study of microquasars and other compact binary sources in the GeV-TeV domain
in recent years has brought new insights into these sources, but probably even
more unanswered questions. In the following years, with the exploitation of {\it
Fermi} data and a new generation of \v{Cerenkov} telescopes, will hopefully
bring the understanding of these sources.

\acknowledgements The authors acknowledge support of the Spanish Ministerio de
Educaci\'on y  Ciencia (MEC) under grant AYA2007-68034-C03-01 and FEDER funds.
V.Z.~was supported by the Spanish Ministerio de Educaci\'on under fellowship FPU
AP2006-00077.

\end{document}